\documentclass[10pt]{article}

\usepackage{arxiv}

\usepackage[utf8]{inputenc} 
\usepackage[T1]{fontenc}    
\usepackage{hyperref}       
\usepackage{url}            
\usepackage{booktabs}       
\usepackage{amsfonts}       
\usepackage{nicefrac}       
\usepackage{microtype}      
\usepackage{lipsum}
\usepackage{graphicx}
\graphicspath{ {./images/} }

\usepackage{subfigure}
\usepackage{algorithm}
\usepackage{algorithmic}
\usepackage{amsmath, amssymb, amsthm}
\usepackage[mathic]{mathtools}

\usepackage{lineno}

\usepackage{thm-restate}
\usepackage[libertine]{newtxmath}
\usepackage[scaled=0.96]{zi4} 

\usepackage{a4wide, dsfont, microtype, xcolor, paralist}

\title{Mixed-coordinate Node-link Visualization for Co-authorship Hypergraph Networks}

\author{
 Mohsen Nafar \\
 Bielefeld University\\
\texttt{mohsen.nafar@uni-bielefeld.de} \\
   \And
 Hamed Azami Zenouzagh \\
  \texttt{aezamihamed@gmail.com} \\
}

\date{\\}

\begin{document}
\maketitle

\begin{abstract}
We present an algorithmic technique for visualizing the co-authorship networks and other networks modeled with hypergraphs (set systems). As more than two researchers can co-author a paper, a direct representation of the interaction of researchers through their joint works cannot be adequately modeled with direct links between the author-nodes. A hypergraph representation of a co-authorship network treats researchers/authors as nodes and papers as hyperedges (sets of authors). The visualization algorithm that we propose is based on one of the well-studied approaches representing both authors and papers as nodes of different classes. Our approach resembles some known ones like anchored maps but introduces some special techniques for optimizing the vertex positioning. The algorithm involves both continuous (force-directed) optimization and discrete optimization for determining the node coordinates. Moreover, one of the novelties of this work is classifying nodes and links using different colors. This usage has a meaningful purpose that helps the viewer to obtain valuable information from the visualization and increases the readability of the layout. The algorithm is tuned to enable the viewer to answer questions specific to co-authorship network studies.
\end{abstract}
\section{Introduction}

The theory of Complex networks is a well-established branch of Computer Science and Mathematics, typically utilizing graphs as models of the real-world systems formed by interacting entities. Those interactions are not always limited to pairwise interactions; thus, graphs, that naturally encode just pairwise interactions, are not always the best way to model these systems. In the case of interactions involving three and more entities, \emph{hypergraphs} seem to be a more adequate mathematical abstraction. Indeed, lately, hypergraphs have found their way into the publication stream in network science \cite{estrada2005complex, antelmi2020analyzing, chodrow2020annotated, roy2015measuring, feng2021hypergraph}. As real-life systems vary significantly by the number of subjects in the groups and other parameters, hypergraphs, that correspond to such systems, are also highly diverse. Thus, it cannot be expected that a single visual presentation or a visualization algorithm could be equally suitable for all the use-cases. Therefore, in the current project we restrict ourselves to specific networks, namely the \emph{co-authorship networks}, in scientific disciplines where typical papers involve no more than a dozen authors. 

The dataset we use in our example is the second-largest connected component of the Mathematics directory on arXiv (https://arxiv.org/), modeled by a hypergraph with \(33\) vertices corresponding to authors, and \(48\) hyperedges corresponding to papers. The sizes of the hyperedges range from \(1\) to \(4\). We believe that our proposed algorithm generally produces competitive results for networks with larger number of smaller hyperedges, i.e. tens to hundreds of nodes with hyperedges of sizes one to ten, not necessarily co-authorship networks.

In this paper, we visualize the hypergraph representing the mentioned dataset. For this specific case, we do not have any constraints for placing the hypergraph's vertices. However, there are several objectives that we seek to reach in our visualization. Among the things we would like to be able to visually track in the network are: authors with the largest number of publications, the most actively collaborating authors, the size of the largest team of authors of a single paper, the most frequent size of a team of authors of a single paper, the most frequent publication number for an author, the connection between different authors and papers, the number of authors who have a specific number of papers, and the appearance of the most active author in different papers. However, we believe that for other networks with similar parameters (modeled with a hypergraph and/or a bipartite graph), our technique is able to address similar questions regarding the relations between different types of entities in the network. 


In this work, each paper-nodes and its adjacent edges are colored according to the number of co-authors of a paper-node. Using different classes of colors for nodes with similar properties assists the viewer in distinguishing some of the graph theoretical properties of the nodes. 



\subsection{Approaches to hypergraph visualization}

As hypergraphs are essentially synonymous to set systems, any set visualization approaches can be applied for visualizing hypergraphs. According to the survey \cite{alsallakh2016state}, set visualization approaches fall into five main categories but in what follows, we combine the first two categories into one.


\subsubsection{Overlays and Euler diagrams}\label{Overlay}

This approach treats hyperedges as closed curves that contains vertices. This class of techniques is similar to that of Euler and Venn diagrams which are vastly being used to visualize sets and their relations and are the most popular members of set visualization techniques \cite{baron1969note}. An example of this approach is \textit{Bubble Sets} visualization in which Collins et al., 2009, \cite{collins2009bubble}, used isocontours to reveal the relation between hyperedges. Similar to those of overlays are Euler diagrams where a set is represented by a closed curve and set relations can be depicted by relation between curves \cite{micallef2014eulerforce}. In 2009, Simonetto et al., \cite{simonetto2009fully}, developed an algorithm that layouts an Euler-like diagram that is suitable for hypergraphs with medium number of hyperedges of medium size. Dinkla et al., 2012, \cite{dinkla2012kelp}, proposed a new visualization approach of the class of Kelp diagrams which they call \textit{Point Set Membership}. They developed this approach to visualize hypergraphs in which the position of their nodes are pre-defined. It seems that this approach can be a good candidate also for hypergraphs with large hyperedges. \textit{SimpleHypergraphs.jl} is a software library that was developed in Julia programming language. It was designed and built by Antelmi et al., 2020, \cite{antelmi2020analyzing}, so as to be used for high-performance computing on hypergraphs. \textit{MetroSets} is the name of an online tool for set systems visualization that is based on a metro map metaphor. This approach was developed and introduced in 2020 by Jacobsen et al., \cite{jacobsen2020metrosets}, that is also compatible to visualize hypergraphs with large hyperedges. Wallinger et al., 2021, \cite{wallinger2021readability}, in a study that targeted to compare LineSets, EulerView, and MetroSets as three of the suitable approaches to visualize medium-sized datasets, quoted that "Our results include statistically significant differences, suggesting that MetroSets performs and scales better".

\subsubsection{Node-link based diagrams:}\label{Node-link}
Both hyperedges and nodes of the hypergraph are represented as two levels of a bipartite graph vertices where an edge between two vertices stands for the set membership in the hypergraph. An algorithm to visualize a bipartite representation of a hypergraph using \textit{anchored maps}, is proposed by Misue, 2006, \cite{misue2006drawing}. Anchored maps techniques are a class of visualization methods in which some of the vertices are restricted to be positioned on some pre-defined places while the rest of vertices have freedom to move; the former vertices are called \textit{anchors} and the latter ones \textit{free vertices}. In this visualization approach, that is the technique for which we have presented an algorithm, one level of vertices are positioned on a circle while the vertices belonging to the other level have freedom to move. A so called \textit{extra-node} representation of the hypergraph was used by Ouvrard et al., 2017, \cite{ouvrard2017networks}, when they were trying to show the improvements it can make for visualising hypergraphs. In fact, it is a visualization of the \textit{star expansion} of the hypergraph using an algorithm called \textit{ForceAtlas2}. This algorithm is a force-directed algorithm that was developed by Jacomy et al., 2014, \cite{jacomy2014forceatlas2}. Another approach is SetCoLa which is a domain-specific language that was designed, created, and contributed by Hoffswell et al., 2018, \cite{hoffswell2018setcola}, to layout graphs using constraints. According to the authors "constraints enable flexible graph layout by combining the ease of automatic layout with customizations for a particular domain". \textit{Py3plex}, a Python library that was implemented and introduced by Škrlj et al., 2019, \cite{vskrlj2019py3plex}. The main purpose of this library was to visualize and analyze networks of multilayer nature. Huang et al., 2020, \cite{huang2020planet}, developed an algorithm called \textit{PLANET} that creates a radial layout of the network. One objective of this algorithm is to minimize the edge crossings while trying to distribute vertices uniformly.  

\begin{figure}
    \centering
    \subfigure{\label{free}
    \includegraphics*[width =.4\linewidth]{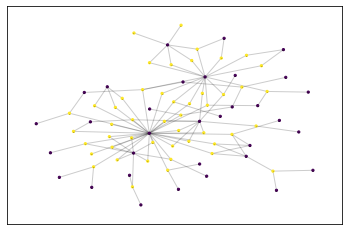}}
    \hspace{3mm}
    \subfigure{\label{anchored}
    \includegraphics*[width =.4\linewidth]{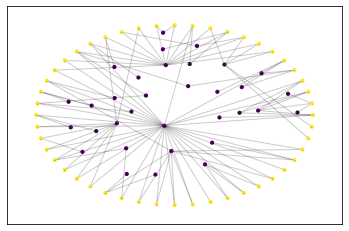}}
    \label{Mixed-coordinate layout}
    \caption{a. Free-coordinate and b. anchored maps layout on our dataset}
\end{figure}

The layout of our dataset was represented as a bipartite graph according to a free coordinate layout (force-directed algorithm) that is the most basic kind of layout is shown in \ref{free}. In this picture, the paper-nodes are colored by yellow and the author-nodes are the purple ones. Figure~\ref{anchored} shows our dataset in anchored map view where the paper-nodes are placed on the oval (pre-defined positions) and the author-nodes are inside the oval. The use of the anchored maps technique improves the picture as it trivializes the distinction of types of the nodes. Our approach is based on the \emph{anchored maps} technique, \cite{misue2006drawing}, which we call a \textit{mixed-coordinate node-link diagram}. Notably, the similarities between the two techniques that use a curve for placing the nodes corresponding to hyperedges, having \textit{semi-fixed coordinate}, and \textit{free coordinate} nodes. A free-coordinate type of a node is restricted within a two-dimensional region on the plane, whereas semi-fixed nodes are restricted to a one-dimensional curve or some reasonably large discrete set of positions. On the other hand, we have \textit{fixed coordinate} nodes, which are not subject to optimization. Moreover, we apply the bundling method in our work. Furthermore, the algorithms developed in the two techniques are totally different.

\subsubsection{Matrix-based diagrams}\label{Matrix-based}

These types of diagrams are basically the visualization of the incidence matrix for a set system. Also some would refer them as matrix metaphors to show sets and their elements. Rows and columns of the matrix represent sets and elements, or vice versa, and entries of the matrix depict set membership relations. \textit{UpSet} is a highly interactive tool that can be used to analyze sets from quantitative point of view. It has various abilities and properties such as showing intersections between sets and group based and query based aggregations \cite{lex2014upset}. As another matrix-based approach, \textit{Bertifier}, a web application, was developed and introduced by Perin et al., 2014, \cite{perin2014revisiting}. Its developers quoted in their paper that their web application uses "Jacques Bertin’s matrix analysis method, whose goal was to “simplify without destroying” by encoding cell values visually and grouping similar rows and columns". Valdivia et al., 2017, \cite{valdivia2017hypenet}, introduced \textit{Hypenet} which they have designed for dynamic hypergraphs visualization. This technique can also be used for pattern and inconsistency detection. \textit{HYPER-MATRIX} is a visual analytic tool for temporal hypergraph model exploration that was presented by Fischer et al., 2020, in \cite{fischer2020visual}. The technique contains various features such as a geometric deep learning model and different interactions to be used as combinations.

\subsubsection{Aggregation-based techniques}\label{Aggregation-based}

In these techniques, some elements can contain multiple data-elements. Approaches of this category strive to demonstrate the cardinality of sets through representing the frequency of elements in set-typed data. \textit{Radial Sets} is an aggregation-based visualization presented by Alsallakh et al., 2013, \cite{alsallakh2013radial}. The authors designed it for visually analysis of a dataset with many elements for their set membership relations. Wang et al., 2015, \cite{wang2015efficient}, developed a software package named \textit{SuperExactTest} that contains a theoretical framework and a visualization technique for multi-set interactions in programming language R. "A Comprehensive Visualization of Set Intersections" is the title of a work by Alsallakh and Ren, 2016, \cite{alsallakh2016powerset}, in which they presented an approach named \textit{PowerSet}. In the paper, it is pointed out that distribution of elements among set intersections can be evaluated. Moreover, it is suitable for exploration and comparison of elements' attributes. \textit{PAOH}, stands for "Parallel Aggregated Ordered Hypergraph", is another visualization approach to layout dynamic hypergraphs. In this technique, vertices are modeled by parallel horizontal bars and vertical lines represent the hyperedges. Valdivia et al., 2019, \cite{valdivia2019analyzing}, fill the entries corresponding to a vertex and a hyperedge that it is belong to by dots.

\section{Proposed visualization technique: mixed-coordinate node-link diagram}\label{worksec}

The approach we are using in the current study belongs to the class of node-link diagrams and deals with the \emph{star expansion} of the hypergraph. This suits our case well as for hypergraphs with relatively small hyperedges (in our sample dataset, the hyperedges cardinalities range from \(1\) to \(4\)), the incidence relations and many graph-theoretic parameters (such as graph distances and centralities) are observable on a node-link diagram. We thus are facing the problem of finding a good layout for the star expansion of a hypergraph. 

In our use-case, it is pretty common that the same group of researchers publish several papers during an extended period of time. In this case, in our hypergraph representation, we might have many hyperedges with high multiplicity (identical as sets). Thus a natural idea is to \emph{bundle} such hyperedges and represent them with a single visual glyph. As an extra visual attribute to encode the hyperedge multiplicity, we use the size of the glyph.

We mentioned that the network contains \(33\) authors and \(48\) papers, which means that before bundling the hyperedges, we would have a bipartite graph with \(81\) nodes in total. After hyperedge bundling, \(48\) paper-nodes are identified and bundled into \(30\) nodes that are displayed. Therefore, the total number of the nodes which was \(81\) is reduced to \(63\). The number of links to be visualized also decreases. 




\begin{algorithm}[htp]
	\caption{Mixed-coordinate layout}
	\label{HBV}
	\noindent\textbf{Input:} A bipartite graph\\
	\noindent\textbf{Output:} Mixed-coordinate layout
	\begin{algorithmic}[1]\baselineskip=14pt\relax
	\STATE \textbf{Initial positioning} 
		\WHILE {(\(E_{T} \ge \text{Threshold} \text{ }\&\text{ } i \le \text{Max}_{iteration}\))}
            \STATE \textbf{Compute repulsion and attraction forces using a modification of force-directed algorithm}
		    \STATE  \textbf{Update positions of non-pendant author-nodes} 
		\ENDWHILE
		\STATE \textbf{Compute crossings}
		\WHILE{(\(i \le \text{iteration}_1\))}
	        \STATE \textbf{Mixed-Discrete-Continuous}
		\ENDWHILE
		\STATE \textbf{positioning pendant author-nodes}
		\RETURN layout.
	\end{algorithmic}
\end{algorithm}
	

We propose restricting the coordinates of the nodes according to the three coordinate types: \emph{free}, \emph{fixed}, and \emph{semi-fixed} coordinates, which we call \emph{mixed-coordinates} (see Introduction). After hyperedge bundling, some author-nodes may become pendant in addition to those that had already been pendant which we call them \textit{pendant author-nodes}. The rest of the author-nodes are called \textit{non-pendant author-nodes}. We restrict the coordinates of paper-nodes to be semi-fixed and we place them on an oval, whereas the coordinates of non-pendant author-nodes are of the free-coordinate type. The coordinates of the pendant author-nodes are of the fixed-coordinate type; we will position these nodes near the paper-node they belong to and outside of the oval. Note that by referring to these nodes as fixed-coordinate nodes, we mean that their positions are fixed relatively to paper-nodes, i.e. only depend on the position of the paper-node they are connected to. The positions of the pendant author-nodes are computed at the end of the algorithm when the positions of all the other nodes are not subject to any change. Furthermore, this new approach enables us to use continuous and discrete optimization methods to compute the layout that simulates the system of forces to find the best configuration.

Moreover, we classify the paper-nodes by the number of their co-authors (cardinality of the original hyperedge) and to make them distinguishable in the picture we encode the number of co-authors of a paper-node with node color. This visual attribute enables the viewer to read out the node degrees with more immediacy, therefore, improves the readability of the layout. To improve the layout further, we color links by the same color as the paper-node they are incident to.

\begin{figure}[H]
\centering
\subfigure[{before improvement}]{\label{Bundled_semi_fixed}
\includegraphics*[width =.4\linewidth]{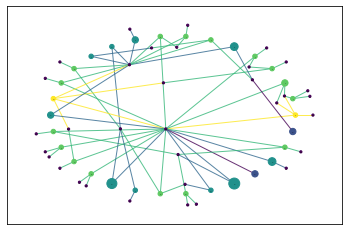}}
\hspace{3mm}
\subfigure[{after improvement}]{\label{near_opt}
\includegraphics*[width=.4\linewidth]{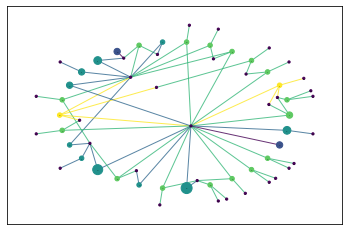}}
\label{Mixed-coordinate layout1}
\caption{Mixed-coordinate layout with/without Mixed-Discrete-Continuous algorithm}
\end{figure}

As for the semi-fixed nodes, the primary vehicle for improving the layout with respect to these nodes is changing the order (permutation) of these nodes on the curve they belong to. This is where the discrete optimization of the algorithm comes in. We have two objectives for the optimization here: minimizing the system's energy and minimizing the number of crossings. A continuous optimization algorithm is used to compute the coordinates of the free nodes. The objective of this part is to minimize the energy of the system. The designed algorithm, which employs the discussed approach, can be found in Algorithm~\ref{HBV}. The result of the Algorithm~\ref{HBV} without lines 7-9 (without discrete part of the algorithm) is shown in \ref{Bundled_semi_fixed}. All the paper-nodes are now placed on the oval, all non-pendant author-nodes are inside of the oval, and every pendant author-node is placed outside of the oval and is connected to the paper-node, which it belongs to, with a short link. The position of every pendant author-node is on an imaginary circle where its center is the corresponding paper-node, but it is important that the author-node be placed on the part of the circle that is outside of the oval that contains the paper-nodes. All author-nodes are purple. In our example, yellow/light-green/dark-green/blue stand for cardinalities \(4,3,2,1\), respectively. The sizes of the paper-nodes correspond to different numbers of hyperedges bundled into the particular paper-node; the more hyperedges are in the bundle, the larger is the corresponding paper-node.

Below you can find a brief description of the procedures of Algorithm~\ref{HBV}. 

\begin{itemize}

    \item  \textbf{Initial positioning.} Position the nodes on two concentric ovals (pendant author-nodes are excluded).
    
    \item \textbf{While loop line 2.} Check the energy decrease (\(E_T\)) against a threshold and check if the number of iterations is below the maximum.




    
    \item \textbf{Compute repulsion and attraction forces using a modification of force-directed algorithm.} Apply an iteration of a simple force-directed algorithm which has the complexity of \(O(n^2)\) per iteration (e.g. \textit{Fruchterman-Reingold algorithm} \cite{fruchterman1991graph}). Note that in this part, forces are computed between three different pairs of nodes: attraction force between a pair of nodes that are connected by an edge, repulsion force between all pairs of non-pendant author-nodes, and repulsion force between pairs of non-pendant author-nodes and paper-nodes. Since all the pendant author-nodes are excluded from this part, we need to modify the force-directed algorithm. However, this part can be done using \textit{Barnes-Hut algorithm} \cite{barnes1986hierarchical} that has complexity of \(O(n\cdot \log n)\) per iteration that needs to be modified for our purpose.

    \item \textbf{Update positions of non-pendant author-nodes.} Update positions of non-pendant author-nodes based on total forces acting on them.

    \item \textbf{Compute crossings.} Compute the number of total and per paper-node crossings that can be accomplished, e.g., in time complexity of \(O(n^2\cdot \log n)\), according to an algorithm in \cite{duque2021counting}.
    
    

    \item \textbf{Mixed-Discrete-Continuous.} This algorithm combines discrete and continuous optimization to improve the layout and is shown in Algorithm~\ref{MDC_alg}.
    
    \item \textbf{Pendant positioning.} Add pendant author-nodes back to the graph and place them close to their paper-nodes outside of the outer oval, no force acts on these nodes in the whole algorithm. To do this part, we consider an imaginary circle around the corresponding paper-node for which the center is the paper-node and place the pendant author-node on its curve in way that it does not lie inside of the oval on which the paper-nodes are placed.
    
\end{itemize}



 \begin{algorithm}[H]
	\caption{Mixed-Discrete-Continuous}
	\label{MDC_alg}
	\noindent\textbf{Input:} Current state of the system\\
	\noindent\textbf{Output:} Improved mixed-coordinate layout using Mixed-Discrete-Continuous algorithm
	\begin{algorithmic}[1]\baselineskip=14pt\relax
        \WHILE{(\(cr^{'} < cr^{*} \text{ }\&\text{ } i \le \text{iterations}_2\))}
            \STATE \textbf{Choose nodes}
            \STATE \textbf{Check number of crossings}
            \IF{(pair is a good candidate)}
                \STATE \textbf{Swap pair}
                \WHILE {(\(E_{T} \ge \text{Threshold} \text{ }\&\text{ } i \le \text{iterations}_3\))}
		            \STATE \textbf{Compute repulsion and attraction forces using a modification of force-directed algorithm like in algorithm~\ref{HBV}} 
		        \ENDWHILE
		    \ENDIF
        \ENDWHILE
	\end{algorithmic}
\end{algorithm}


As previously discussed, the main idea of Mixed-Discrete-Continuous algorithm is to add another objective (minimizing the number of crossings) to the problem and solve it by combining discrete and continuous optimization. The reader may compare the quality of the layouts of \ref{Bundled_semi_fixed} (before improvement) with \ref{near_opt} that is the result of the improvement made by our algorithm. Note the decrease in the number of crossings when comparing the two layouts. We will briefly describe Algorithm~\ref{MDC_alg} in the following.

\begin{itemize}
    \item \textbf{While loop line 1.} Makes sure that the loop runs while the crossing number has not decreased (\(cr^{'}\) refers to the number of crossings before the loop and \(cr^{*}\) refers to the number of crossings after one iteration) \emph{and} the limit on the maximum number of iterations has not been reached. 
    
    \item \textbf{Choose nodes.} Select two candidates from paper-nodes to swap them.
    
    \item \textbf{Check number of crossings.} Check to see whether swapping the chosen paper-nodes will reduce the number of crossings, we can simply do it in linear time (\(O(n)\)); since we already know the number of crossings every paper-node participates in, we need to compute the number of crossings that these nodes cause in case we replace them with each other.
    
    \item \textbf{Swap pair.} Exchange the position of the chosen paper-nodes. 


    
\end{itemize}





\section{Discussion and results}

Assigning different colors to paper-nodes with different cardinalities makes it easier for the user to visually approximate the number of the papers with certain number of authors, which was one of the aims we introduced in introduction of this work as an objective. Since the color of every edge is the same as the paper-node it is incident to, by looking at an author-node the user can easily track what the sizes of the collaboration groups (papers) that an author participated in are, which is another aim that we set for our technique. Moreover, looking at author-nodes, the user can visually notice those that have published the largest number of papers. Although bundling the paper-nodes decreases the degree of the author-nodes it is connected to, different size of the bundled paper-nodes that depends on the number of the articles in the bundle solves this problem, therefore, the user is able to find the most active authors. Furthermore, the user is able to understand the relation between any two authors by tracking their incident links and since the colors of the links have a particular meaning the user does not need to track the edges that have different colors (e.g. two authors under consideration do not have links with the same color, then they surly have no common paper). In addition to what we discussed so far, the user is able to find the most frequent size of a team of authors of a single paper by just visually tracking the number of paper-nodes that have the same color and comparing for different colors. Therefore, our technique enables the user not only to see the relations between vertices of the original hypergraph (in our case author-nodes), but also the relations between hyperedges and vertices of the original hypergraph (paper-nodes and author-nodes). We believe that the information that our technique is able to reveal can not be revealed by other visualization approaches, unless a combination of approaches get in use. The color coding of the paper-nodes and links also helps in situations when a node is placed close to a link which is not incident to. 

\begin{figure}[H]
\centering
\subfigure[{SimpleHypergraphs.jl}]{\label{SimpleHypergraphs.jl}
\includegraphics*[width =.4\linewidth]{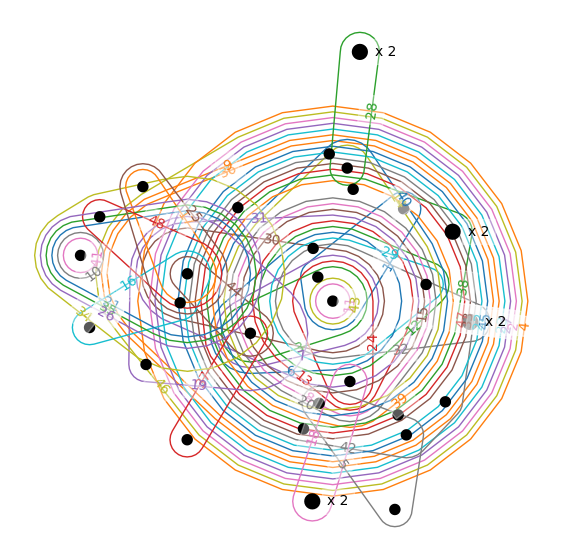}}
\hspace{3mm}
\subfigure[{ForceAtlas2}]{\label{ForceAtlas2}
\includegraphics*[width=.4\linewidth]{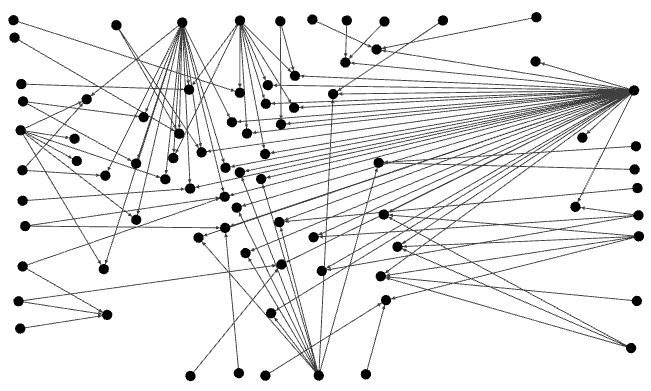}}
\hspace{3mm}
\subfigure[{Bertifier Matrix}]{\label{Bertifier Matrix}
\includegraphics*[width=.4\linewidth]{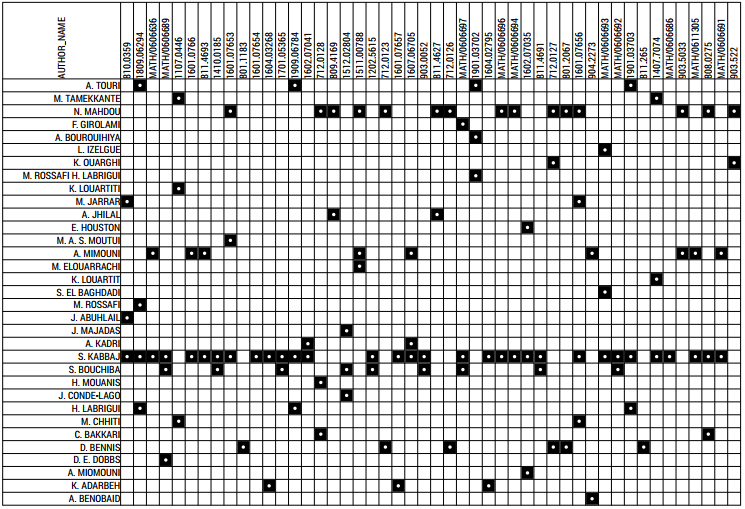}}
\hspace{3mm}
\subfigure[{PAOH}]{\label{PAOH}
\includegraphics*[width=.5\linewidth]{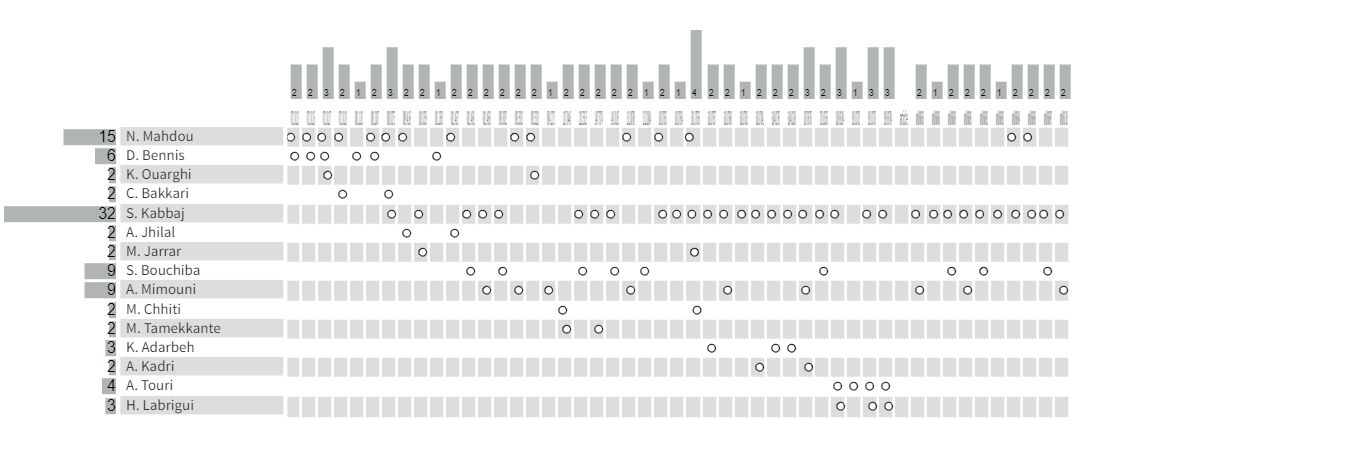}}
\caption{Four visualization methods output for a data set with 33 authors and 48 articles}
\label{math_second_comp}
\end{figure}

In current section, we compare our technique with four different visualization methods which belong to different categories of hypergraph visualization techniques. Some of them are the-states-of-the-art in the visualization literature, i.e. SimpleHypergraphs.jl, ForceAtlas2, Bertifier, and Parallel Aggregated Ordered Hypergraph Visualization (PAOH). Figure \ref{math_second_comp} shows the output of these visualization techniques on the dataset we used in this paper. We evaluated the methods with each other according to these questions: 
Q1. Which authors do have the largest number of publications? Q2. Who is the most actively collaborating author? Q3. Who are the most actively collaborating team of authors? Q4. What is the size of the largest team of authors of a single paper? Q5. What is the most frequent size of a team of authors of a single paper? Q6. What is the most frequent publication number for an author? Q7. What are the connection between different authors and papers? Q8. What is the number of authors who have a specific number of papers? Q9. what is the appearance of the most active author in different papers?

We summarized the results of the comparison of the techniques in answering the questions in table \ref{compare our method}. In this table, a method obtains check-mark for a question, if a viewer answers the question faster by only looking at the layout and without doing any calculation except comparing values. In other words, additional calculations such as summation of values are not allowed and leads to cross-sign. If the number of an author's publication is much greater than the others, a viewer can answer Q1 by using all of these five methods. However, if the difference is small, PAOH and our method act better than the others. The reason that PAOH and our technique are able to lead the viewer to answer the Q1 faster is that we encode the degree of paper-nodes with specific colors and PAOH possesses additional information in row and column. According to our method properties such as nodes positioning and using colors in a meaningful way, the method enables the viewer to answer Q2 and Q3 which cannot be answered by using other methods. SimpleHypergraph.jl and ForceAtlas2 failed in finding the size of the largest team of authors of a single paper. In Bertin's matrix method, the viewer should add up the values of each column to obtain the number of authors for each paper and then answers Q4. By comparing the additional information in PAOH, a viewer can answer Q4. Finding the largest number is even easier with our layout because the problem is changed into checking the colors assigned to specific degree of paper-nodes which is a very easy visual task. For Q5, Q6, and Q8, the description is the same as that one which was stated for Q4. In these cases, the viewer focuses on the repetition number of specific colors (as they reveal very specific information about the paper-nodes). In our technique, the connection between authors and articles are shown by meaningful colors. As a result, these connections are clear using our method which the others lack this property. Additional information of PAOH and the color-using property of our method enables the viewer to answer Q9. Although, Mixed-coordinate Node-link Visualization method visualizes complicated networks, the trustworthy of the method are not as good as that ones which are obtained for networks with larger number of smaller hyperedges.

\begin{table}[H]
\centering
\begin{tabular}{|p{0.4cm}|p{3.1cm}|p{2cm}|p{2.3cm}|p{1.5cm}|p{2.1cm}|} 
 \hline
 Qs  & \hfil SimpleHypergraphs.jl & \hfil ForceAtlas2 & \hfil Bertin’s matrix & \hfil PAOH & \hfil Our technique\\ [0.5ex] 
 \hline\hline
 1  & \hfil \checkmark & \hfil \checkmark & \hfil \checkmark & \hfil \checkmark & \hfil \checkmark\\
 \hline
 2  & \hfil $\times$ & \hfil $\times$ & \hfil $\times$ & \hfil $\times$ & \hfil \checkmark\\ 
 \hline
  3  & \hfil $\times$ & \hfil $\times$ & \hfil $\times$ & \hfil $\times$ & \hfil \checkmark\\ 
 \hline
  4  & \hfil $\times$ & \hfil $\times$ & \hfil $\times$ & \hfil \checkmark & \hfil \checkmark\\ 
 \hline
 5  & \hfil $\times$ & \hfil $\times$ & \hfil $\times$ & \hfil \checkmark & \hfil \checkmark\\ 
 \hline
  6  & \hfil $\times$ & \hfil $\times$ & \hfil $\times$ & \hfil \checkmark & \hfil \checkmark\\  
 \hline
  7  & \hfil $\times$ & \hfil $\times$ & \hfil $\times$ & \hfil $\times$ & \hfil \checkmark\\ 
 \hline
  8  & \hfil $\times$ & \hfil $\times$ & \hfil $\times$ & \hfil \checkmark & \hfil \checkmark\\  
 \hline
  9   & \hfil $\times$ & \hfil $\times$ & \hfil $\times$ & \hfil \checkmark & \hfil \checkmark\\ [1ex] 
 \hline 
 
\end{tabular}
\caption{Comparing our method with 4 different visualizations methods}
\label{compare our method}
\end{table}

\section{Conclusion}

The algorithm that we have proposed has several novelties that benefits the readability of the resulting layout. We combined both freely positioned nodes and nodes attached to the fixed circle, thus providing clear visual difference between the nodes corresponding to the authors and the papers. Moreover, we employ node bundling and special positioning of the pendant nodes to maximally de-clutter the visualization in center of the canvas. We use color coding to make it easy to visually estimate node degrees, also additional use of color coding on the links allows for estimation of the degrees of the paper-nodes that neighbour a given author-node. Furthermore, we have combined continuous and discrete optimization in a single algorithm for best visual results. Below we outline some areas for improvement on the proposed technique and topics for further investigation.

Firstly, in Algorithm~\ref{MDC_alg}, one can have a multitude of strategies for choosing the two paper-nodes as the candidates for swapping in the discrete optimization stage of the algorithm. One such strategy is to choose the nodes one after another, and while doing it, we can have different criteria for these choices. For instance, we can choose the first paper-node randomly or greedily; take the paper-node with the highest energy or the paper-node with the highest share in the crossings. Note that we can make multiple copies of the system after choosing the first paper-node and compute different cases for different candidate choices of the second paper-node in parallel. Moreover, the choice of the two paper-nodes can be dependent on each other or independent. Furthermore, it is possible to choose a pair of paper-nodes at once rather than choose the two vertices in consecutive steps. However, currently in this part of the algorithm we choose the two paper-nodes consecutively, randomly, independently of each other, and according to their share in the number of crossings. The other mentioned cases remain for further investigation. 

Secondly, the cardinality of the discrete set of positions for placing semi-fixed coordinate nodes (paper-nodes) is equal to the number of paper-nodes after bundling. This means that paper-nodes are equally distanced on the oval. We believe that the layout may benefit from uneven distribution of the paper-nodes on the oval in some cases, although we do not have the corresponding implementation for it right now. 

Lastly, the approach that we have presented is adequate for the visualization of hypergraphs sharing similar distributions of node degrees and hyperedge cardinalities as ours. The question remaining for further investigation is how far our approach scales for other types of networks. Moreover, in our approach, we have a circular area and a circumference on which the nodes of some particular type are placed. We are currently investigating a novel visualization type where nodes corresponding to hyperedges of different cardinalities are placed on different concentric circles surrounding the area of author-nodes. The corresponding work, which is an extension of the current paper, is under preparation.


\section*{Acknowledgements}

We gratefully acknowledge the assistance of professor Elena Bazanova for her helpful contributions in the editing phase of the paper.

\bibliography{./biblio}
\end{document}